# Adaptation through stochastic switching into transient mutators in finite asexual populations.

Muyoung Heo, Louis Kang and Eugene I. Shakhnovich

Department of Chemistry and Chemical Biology, Harvard University,
12 Oxford St, Cambridge, MA

# Abstract

The importance of mutator clones in the adaptive evolution of asexual populations is not fully understood. Here we address this problem by using an ab initio microscopic model of living cells, whose fitness is derived directly from their genomes using a biophysically realistic model of protein folding and interactions in the cytoplasm. The model organisms contain replication controlling genes (DCGs) and genes modeling the mismatch repair (MMR) complexes. We find that adaptation occurs through the transient fixation of a mutator phenotype, regardless of particular perturbations in the fitness landscape. The microscopic pathway of adaptation follows a well-defined set of events: stochastic switching to the mutator phenotype first, then mutation in the MMR complex that hitchhikes with a beneficial mutation in the DCGs, and finally a compensating mutation in the MMR complex returning the population to a non-mutator phenotype. Similarity of these results to reported adaptation events points out to robust universal physical principles of evolutionary adaptation.

Mutators are clones with high mutation rates, and they play a major role in adaptation to a new environment (*1-4*). Natural populations exhibit a broad range of mutator allele frequencies which are relatively higher than expected (*5-9*). Because mutators can rapidly produce beneficial mutations, they can get fixed in the population by hitchhiking (*10-12*). However, they also burden the population with deleterious mutations which eventually outnumber beneficial ones, and thus mutation rate diminishes to a minimum in well-adapted populations (*13*). Despite many studies on mutator clones which have been carried out in natural isolates (*5-8*), laboratory bacterial strains (*14, 15*), and computer simulations (*1, 16*), the microscopic mechanism that underpins the emergence, fixation, and disappearance of mutator clones still remains unknown.

Here, we developed a microscopic ab initio model of organisms to study evolutionary dynamics of mutator clones in a finite asexual population. Each organism carries five genes with corresponding protein products. The first three genes are housekeeping genes responsible for cell growth and division, (replication controlling genes or RCGs) and products of genes 4 and 5 dimerize to form a mismatch repair (MMR) complex – mimicking mutS systems in bacteria which are active in vivo as tetramers (dimers of dimers)(17) . The three RCGs form a simplest functional PPI network where protein 1 functions in isolation and proteins 2 and 3 must forma functional dimeric complex. The model with three RCGs was used in our recent study (18) where it was shown that this model is a minimal one which takes into account protein function (Protein-Protein Interactions) and is capable of reproducing rich biology of evolution of mutation rates.

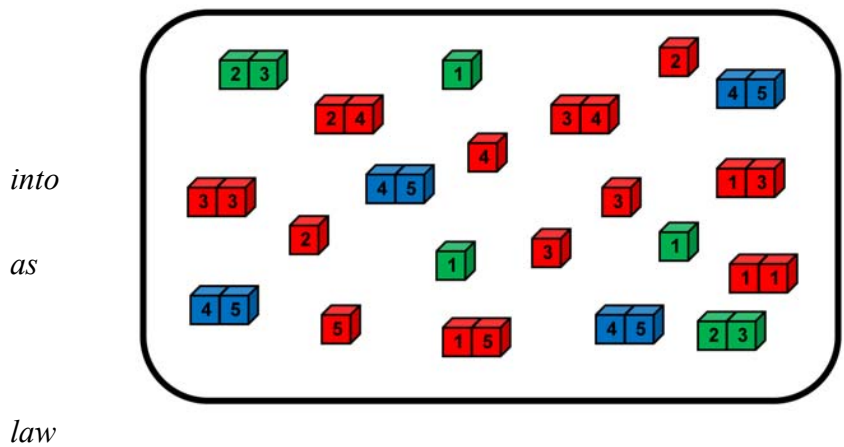

**Figure 1. *Schematic diagram of the model.*** *A model organism has 5 genes, which are expressed into multiple copies of model proteins. Proteins can stay as monomers or form dimers whose concentrations are determined by interaction energies among them and law of mass action equations. Green cubes represent proteins in their functional states that increase an organism's birth rate. Blue cubes represent functional MMR dimers: When their concentration is above the threshold value, an organism is in a non-mutator state. Red cubes represent proteins in their non-functional states.*

The fitness, i.e. the growth rate, $b$ of an organism is determined by the monomer concentration of the first protein and dimer concentration of the second and third proteins (18)

$$b = b_0 \frac{F_1 \cdot F_{23} \cdot P_{int}^{23} \cdot \prod_{i=1}^{3} P_{nat}^i}{1 + \alpha \left( \sum_{i=1}^{5} C_i - C_0 \right)^2}, \quad (1)$$

where $b_0$ is a base growth rate, $F_i$ is concentration of monomeric protein $i$ and $F_{ij}$ is concentration of heterodimer complex between protein $i$ and $j$ in all possible binding configurations. $P_{int}^{23}$ is the Boltzmann probability of binding between protein *2* and *3* in a native, functional binding configuration whose binding energy has the lowest value of all possible mutual configurations and $P_{nat}^i$ is stability (Boltzmann probability to be in the native state) for the protein product of gene $i$. $C_i$ is total production level for protein $i$, $C_0$ is an optimal total production level for all proteins in a cell, and $\alpha$ is a control coefficient which sets the range of allowed deviations from optimal production levels. The protein products of fourth and fifth genes determine mutation rate by acting as components of an MMR system of DNA replication (*2, 5, 15*). The fidelity of an organism's DNA replication is controlled by its dimer concentration of the proteins 4 and 5 bound together as a structurally well-defined specific functional MMR complex: $G_{45} = F_{45} \cdot P_{int}^{45}$. A non-mutator can be transformed into a mutator whose mutation rate is 500-fold higher when $G_{45}$ drops below a threshold value. Reversely, the rise of $G_{45}$ above the threshold value brings a mutator back to a non-mutator state. The $C_i$s can fluctuate, reflecting noise in protein copy numbers in living cells, and $F_i$ and $F_{ij}$ are exactly calculated for a given set of $C_i$ by solving equations of the Law of Mass Action (LMA) (*19*). Thus, a mutator can emerge by a drop in MMR gene expression levels, which affects $F_{45}$, or by mutations of the MMR genes that disfavor complex formation, which affect both $F_{45}$ and $P_{int}^{45}$. (See Figure 1 and Methods for illustration and details)

Fig. 2 shows the evolution of populations which undergo adaptation. A mutator hitchhiked with mutations in the DCGs increasing the fitness from $b$=0.005 to 0.05, stayed fixed in the population until $b$ reached 0.5 and then disappeared (See Fig 2A and Supplementary Figure 1). Further adaptation from $b$=0.5 to 0.6 did not involve mutator fixation once the population obtained high fitness. After that, new mutators emerged continuously until the end of simulation, but they could not get fixed again. We prepared another unfavorable stressor to the population by increasing environmental temperature from T=0.85 to 1.0 at t=20000 (Fig. 2B) and checked whether a mutator was always involved in adaptation into such a new environment.

The temperature jump initially plunged the fitness down threefold to $b=0.14$ and at the same time a mutator got fixed again.

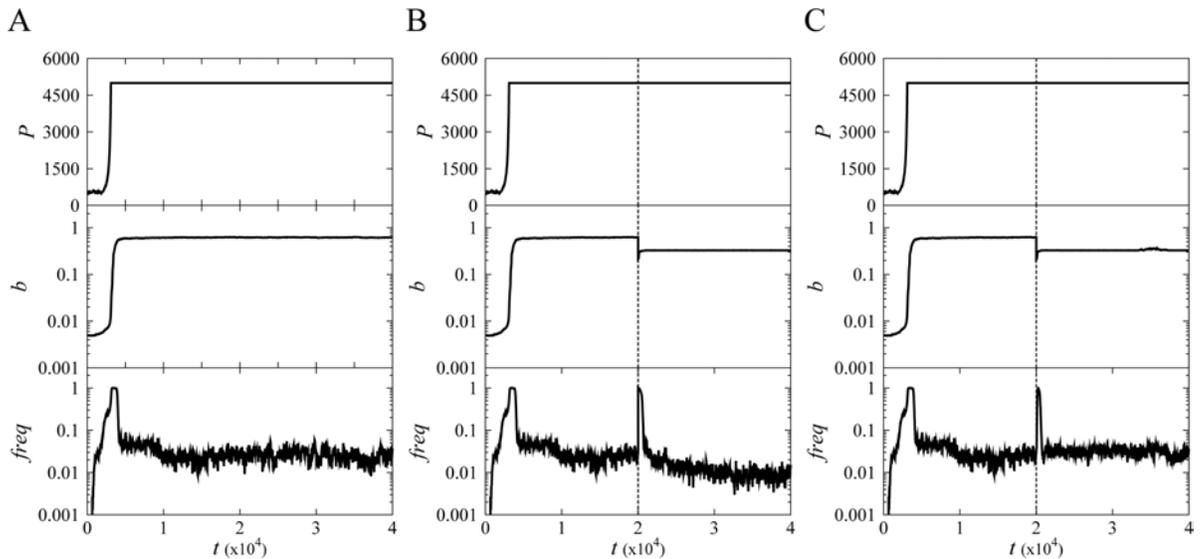

**Figure 2.** *Population dynamics of the model.* *Panels show population (P), mean birth rate (b), and frequency of mutator allele (freq) in the population as function of time (t). (A) To escape from an initially low fitness state, a mutator must get fixed and generate beneficial mutations that provide higher b. Once the organisms reach a high fitness state, genomic stability and the non-mutator phenotype are preferred. (B) The blue line in each panel at $t=20000$ marks the time of temperature increase. The temperature jump causes a drop in b and another mutator fixation event. (C) The blue line in each panel marks the time of birth rate decrease. This drop creates selective pressure that prefers mutators, which soon get fixed in the population. This mutator fixation allows the population to find a beneficial mutation that enhances birth rate, and then non-mutators are preferred after adaptation. These simulations use the same pseudorandom number generator seed.*

Experimental studies of a bacterial population grown in a laboratory chemostat demonstrated fixation of an initially abundant mutator, but otherwise when mutators appeared at low frequency, they were purged from the population (20). Later, de Visser *et al.* showed that a mutator has selective advantage in a small population and the initial "adaptedness" of the population determines the role of mutators in enhancing fitness (2). We derived two hypotheses from these two experimental results. One is that a high level of fitness can exclude mutator fixation by limiting the occurrence of beneficial mutations with which mutators hitchhike. The other is that a rapidly growing population can discard mutators by random drift. Our microscopic model provides a direct opportunity to test these hypotheses as it allows to exactly evaluate the effect of point mutations on organismal fitness. To this end, we first examined whether maladapted populations are more likely to produce beneficial mutations than adapted ones. We investigated the distribution of relative fitness change upon single random point mutations in our microscopic model. We selected dominant clones at various time points and computed the relative fitness changes according to Eq.1 (assuming all concentrations C unchanged) of 1000 mutant genomes, each of which differs from the dominant clone by a single mutation. Fig 3 shows histograms counting the number of mutants with a certain relative fitness change. In Fig 3A, at t=2500 the long tail in the beneficial region for the maladapted clones demonstrates a mutational bias towards higher probability of beneficial mutations, and such tail disappears after adaptation at t=5500 suggesting that beneficial mutations become less available when population adapts. We also examined the perturbation of mutational bias upon temperature jump and adaptation to the high temperature by comparing the distribution of the relative fitness changes among 3 dominant clones, each from a different population. The distributions in red, blue and green in Fig 3B correspond to the populations in Fig 2A at t=20,000 (an adapted population at low temperature), in Fig 2B at t=20,000 (a maladapted population at high temperatures) and in Fig 2B at t=26,000 (an adapted population at high temperature). We found that neither the environmental change by temperature jump (between red and blue distributions) nor the adaptation to high temperature (between blue and green) could perturb the balance between beneficial and detrimental mutations.

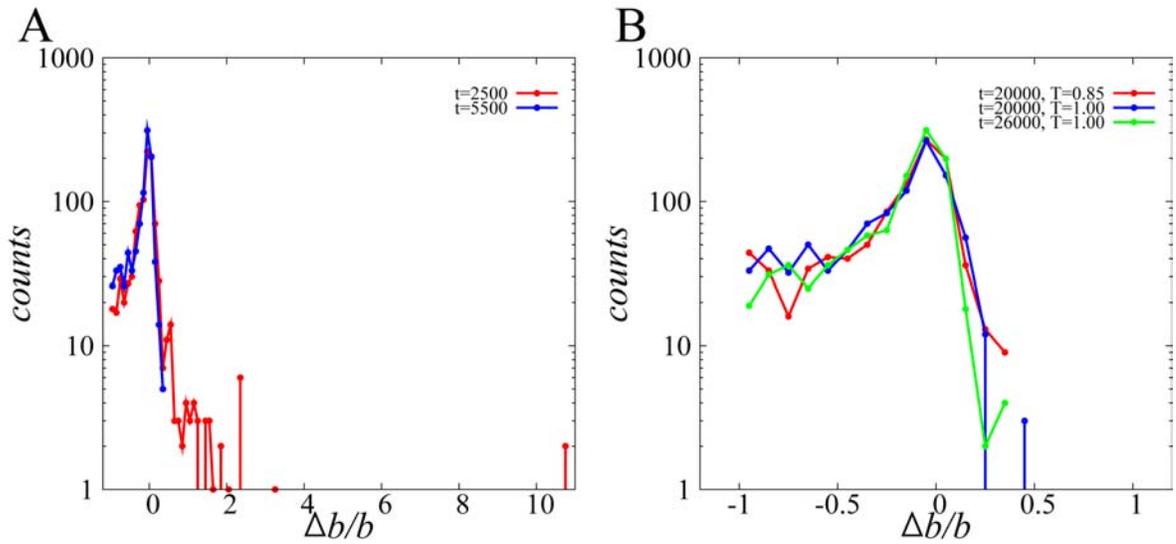

**Figure 3.** *Distributions of relative fitness changes upon a single mutation. Organisms were created using the genomes of the dominant clones at certain time points in simulation (B) of Figure 2 and using the mean protein concentrations $C_i$ at those time points as the organism's $C_i$ values. The organisms' genomes were then each subjected to 1000 separate non-synonymous random point mutations and the birth rate was calculated for each of the 1000 mutants per created organism. $\Delta b/b$ gives the fractional difference in the birth rate between each original created organism and its corresponding 1000 mutants. The histogram counts the number of organisms at a certain $\Delta b/b$. (A) A distinct change in single-mutation fitness distribution can be seen before and after the first mutator fixation. Organisms at t=5500 with higher birth rate show less tendency to further increase birth rate than organisms at t=2500 with lower birth rate. (B) However, no distinct change in fitness distribution can be seen among the dominant clone*

*immediately before the temperature increase, immediately after the temperature increase, and after adaptation to the higher temperature. Thus, although higher temperature lowers the birth rate, it does not lead to a greater genomic bias towards more beneficial mutations.*

We found therefore that random genomes could experience many beneficial mutations, but once the population was adapted to the environment, i.e. reached a local fitness peak, temperature jump could neither distort nor bias the fitness landscape to generate additional beneficial mutations. However, as shown in Figs 2B and 3B, mutator clones became fixed regardless of the mutational bias in the fitness landscape and the availability of beneficial mutations. Next, we checked if the level of fitness, i.e. the growth rate itself, affects the fixation of mutators by random drift. We simulated a starvation condition by abruptly dropping the growth rates of all organisms by 3-fold (by decreasing the value of $b_0$ in Eq.(1) at $t$=20,000). Fig 2C shows that simulated starvation condition caused another transient fixation of a mutator clone. For a control, we carried out simulations which started with a high initial level of fitness ($b$=0.6, which is 120-fold higher than normal model) and surprisingly, no mutators became fixed in this condition in the course of adaptation (Supplementary Figure 2). From both de Visser's experiment (*2*) and our simulations, we can reason about the condition of mutator fixation in the following way. High fitness implies faster organism reproduction and greater removal of excess organisms in an environment that can only support a finite population. Unless they generate beneficial mutations, mutators introduced to the system at low frequency are highly susceptible to removal by random drift because their reproduction is limited due to some loss of fitness caused by high mutation rate (see Supplementary Figure 3). If fitness is low, then a mutator can wait until it generates a beneficial mutation and gets fixed by hitchhiking. But otherwise, it may be quickly washed out by random drift before generating a beneficial mutation. To test this reasoning, we performed another control simulation with fixed fitness (rather than fitness determined by Eq.1) and found that the level of fitness indeed determines the frequency of mutators through random drift (see Supplementary Text and Supplementary Figure 3).

Now we turn to the microscopic anatomy of an adaptation event. The dynamics of microscopic variables such as protein production levels $C_i$, concentration of functional MMR complexes $G_{45}$, and Boltzmann probability $P_{\text{int}}^{45}$ to form functional MMR complexes are shown in Fig. 4. These data provide insights into molecular mechanisms underlying the emergence, fixation and disappearance of mutator clones. Two factors are potentially responsible for emergence of mutators: stochastic switching through fluctuation of production levels $C_i$ and mutations changing the stability of the MMR complex $P_{\text{int}}^{45}$. The initial set of $C_i$ converged to a more optimal distribution by reallocating resources for better fitness: The concentrations of housekeeping proteins ($C_1, C_2$, and $C_3$) increased, while concentrations of the MMR complex proteins, ($C_4$ and $C_5$) decreased. These phenotypic changes in protein production levels were the primary factors causing the rise of mutators at early stages of adaptation. Similar parallel changes in gene expression pattern were also reported in long-term evolutionary experiments (21, 22).

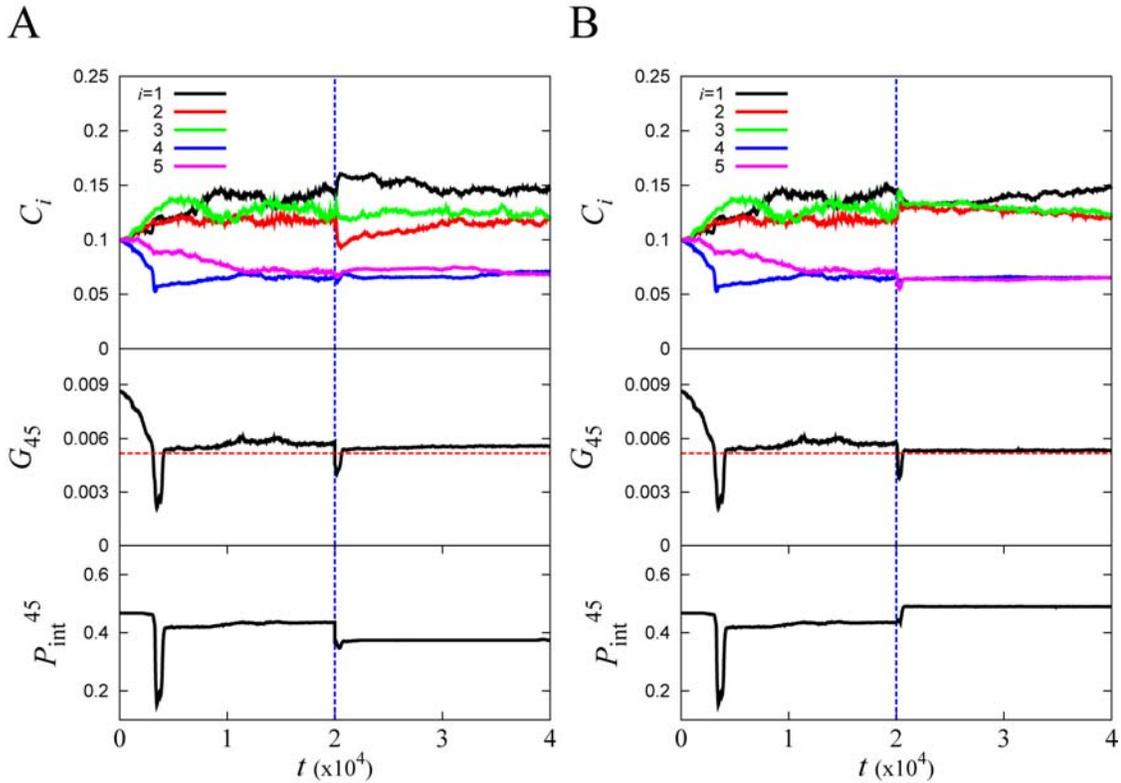

**Figure 4.** *Microscopic anatomy of adaptation events.* Panels present mean concentrations ($C_i$) of each protein product, mean heterodimer concentration of proteins 4 and 5 which constitute the functional MMR complex ($G_{45} = F_{45} \cdot P_{int}^{45}$), and mean heterodimer stability of the functional complex ($P_{int}^{45}$) as function of time (t) corresponding to population dynamics of adaptation processes shown in Fig. 1B (A) and in Fig. 1C (B). The red line in the $G_{45}$ panel is set at $G_{45}^c$, the critical heterodimer concentration below which an organism becomes a mutator. Mutator clones at t=20000 were induced by loss of $P_{int}^{45}$ by temperature jump (A) and by stochastic switching (B), but both of them disappeared after adaptation as new mutations restored the MMR system by enhancing $P_{int}^{45}$.

After high mutation rate was reached through protein concentration fluctuation, a mutation which decreased the stability of the MMR complex was quickly found and fixed in population along with the mutator phenotype (Fig.4). However, upon adaptation, a compensatory mutation occurred to increase $P_{int}^{45}$ and eliminate mutators, and the functional MMR complex concentration, $G_{45}$, then stayed just above the threshold value because no further selective pressure was exerted on it. The close-to-threshold level of $G_{45}$ can enable the population to generate mutator clones quickly by stochastic switching like the second adaptation events at $t$=20000 in Figs. 2B and 2C. In order to determine precisely the microscopic causes of phenotypic switches between mutators and non-mutators, we traced all transitions between them in Fig. 5, and checked more closely whether stochastic switching or mutation caused the transformations. Green lines in all panels of Fig. 5 indicated that a mutator phenotype was switched on or off by variation of protein production levels $C_i$. All mutators in the bottom panels of Fig. 4, except in the temperature jump case, initially emerged from stochastic variation of protein production, i.e. they represented phenotypic switches. The temperature jump effect is the only exception because high temperature disrupted the heterodimer binding stability of the MMR complex.

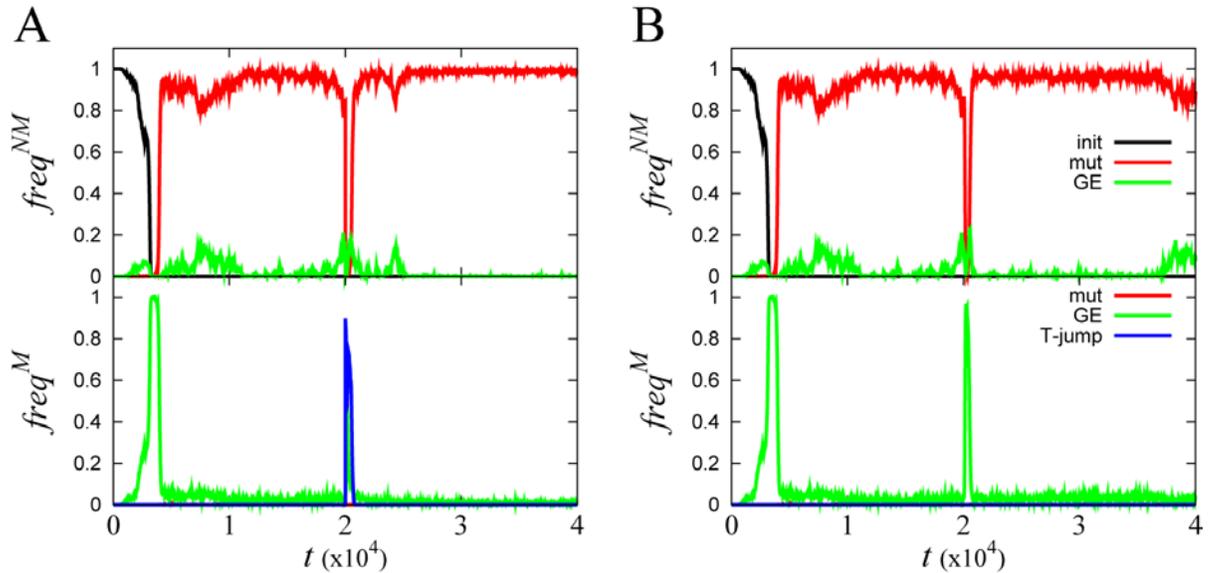

**Figure 5.** *__The rise and fall of non-mutator and mutator clones.__ The frequencies of non-mutator (freq$^{NM}$) and mutator (freq$^M$) clones that arise from different evolutionary forces are shown as function of time (t). Each color represents the cause of the organisms' most recent transformations from non-mutator to mutator, or vice versa. Red represents genotypic mutation, green represents stochastic phenotype switching, and blue represents a temperature jump. The black line is the original non-mutator population. Panels A and B respectively correspond to population dynamics in Fig. 2B and Fig. 2C.*

Why did mutators emerge through stochastic switching rather than a genotypic change (mutation)? To address this question we studied adaptation at various rates $r$ of stochastic fluctuation of protein production, from $r=10^{-2}$ to $10^{-3}$, $10^{-4}$, and $r=0$ – the case where no fluctuations of protein production were allowed (Fig. 6; see Supplementary text for details of definition of fluctuation rates $r$). Deceleration of the fluctuation rate delayed the fixation of mutators, and furthermore, no mutators (and, strikingly, adaptation) were observed when $r=0$. In fact, the transitions between mutators and non-mutators always occurred in a specific microscopic order. First, a stochastic phenotype switching (encouraged by resource reallocation or close-to-threshold $G_{45}$) created mutators in the population. These mutators then generated destabilizing mutations in MMR complex and beneficial mutations in genes 1-3 and hitchhiked with them to fixation. Subsequently, mutators disappeared through the fixation of mutations which provided a more stable MMR system.

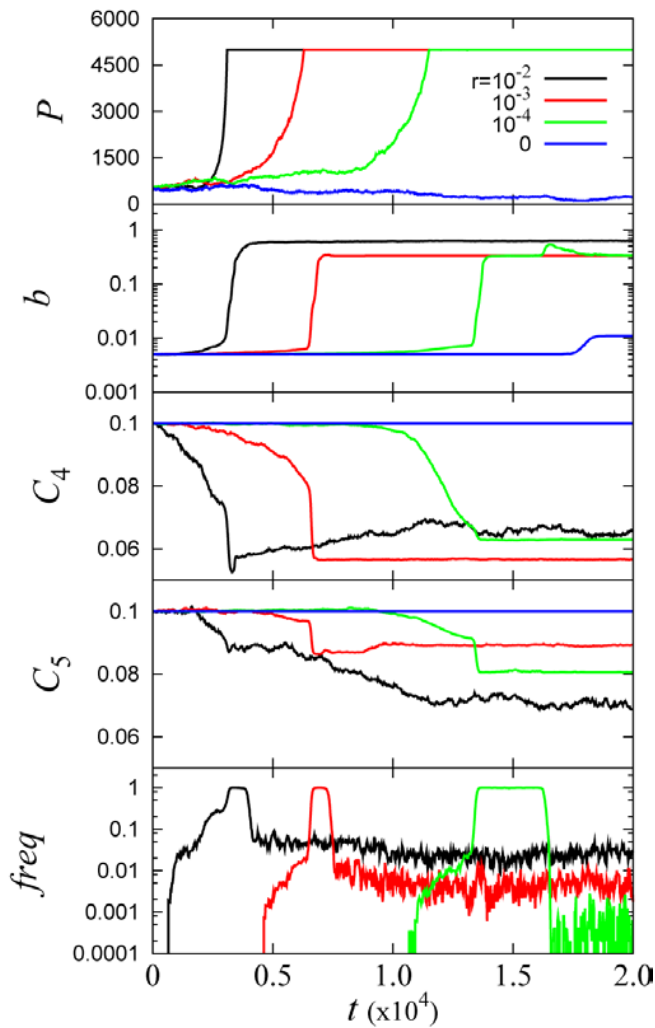

**Figure 6. *Stochastic switching to mutators.*** *Population (P), mean birth rate (b), mean total concentration of protein 4 ($C_4$) and 5 ($C_5$), and frequency of mutator allele (freq) in the population are plotted as a function of time (t). The lines represent simulations at various fluctuation rates: $r=10^{-2}$ (black), $10^{-3}$ (red), $10^{-4}$ (green) and 0 (blue)(see SI for details of definition of fluctuation rates r). All non-zero traces showed decreased gene expression levels ($C_4$, $C_5$) of MMR complex proteins. The concentration in black was vacillating due to the high fluctuation rate, while the concentrations in red and green seemed locked into optimal values. Without stochastic switching (blue), no mutators could arise and the adaptation does not occur.*

These findings provide a general framework to understand the physical mechanisms of adaptation under a stressful environment such as antibiotics, starvation, or heat-shock. In particular they point out to the fundamental and universal role of the stationary phase in bacterial adaptation. Bacteria are known to have evolved with specific regulators which adjust their mutation rates to cope with environmental stresses (3, 23). But their basic strategy against stress is to switch into a stationary phase in which they can persist with a limited growth rate (24) and adapt through transient increase of their mutation rate (23, 25). Experimental data shows that entrance into a stationary state is marked by a sharp drop in the expression of MMR genes (26, 27). This drop, possibly occurring due to resource reallocation, is predicted by our model as a universal physical mechanism to enter the adaptation stage through a quick transient fixation of mutator clones. A low-fitness state prevents these transient mutators from being eliminated by random drift long enough to find mutations destabilizing the MMR complex. Such genetic change prevents the mutators from being eliminated by reverse fluctuations of MMR gene expression levels and allows them to become fixed. Once the stationary state bacteria find beneficial mutations with greater fitness, they can revert back to non-mutators.

This model captures realistically, without ad hoc assumptions, the change of mutation supply upon adaptation and it shows that mutator can be fixed in a new environment that does not significantly perturb the fitness landscape. (see Fig. 2B, 2C and Supplementary Figure 1). The other novel aspect of our model is that the biological functions of an organism are derived from a simple genetic network, where gene expression levels and their fluctuations are taken into account explicitly. The impact of protein production levels on cell fitness is evaluated in the model through the rules of physical chemistry which govern interactions between proteins in the cytoplasm of model cells. We found that noise in gene expression plays a key role in adaptation by providing an initial supply of mutators by affecting the transient stability of the functional MMR gene products. Strikingly, the model organisms lost their ability to adapt through a transient mutator phenotype when gene expression noise was eliminated (see Fig.6).

A key component of the *E. coli* MMR system, *mut*S, is efficient in the tetrameric form (17). It is noteworthy that at conditions of exponential growth, the concentration of *mut*S dimers is close to the threshold of the dimer-tetramer equilibrium transition,(17, 26) in direct support of our finding that the expression and stability of the functional MMR complex in adapted populations our model is maintained very close to the mutator threshold (Fig.4). The proximity of the concentration of the MMR components to the critical threshold explains the persistent presence of a small proportion (1-10%) of mutators in the populations observed in our simulations (Fig.2) and in experiment. (*4*) (*3*). They are immediately available when the need for mutators in adaptation arises (*14*). Our simulations provided a detailed insight, on a genomic and proteomic level, into the set of events which lead to adaptation. First, a mutator is supplied by a fluctuational decrease in MMR protein production, leading to a steep drop in the number of functional multimeric complexes. Mutator cells then acquire mutations which destabilize the MMR protein complexes making them more prone to persistence, but mutators can get fixed in the population only when fitness is limited. Next, the mutator phenotype reaches 100% fixation via hitchhiking with a beneficial mutation in the DCGs. Finally, a compensating mutation in an MMR gene arrives and a non-mutator strain, adapted to new conditions, gets fixed in the population, concluding the adaptation event. The important role of recurrent losses and reacquisition of MMR gene functions was highlighted in the study by Denamur et al. (28) who found that phylogeny of the MMR genes in *E. coli* is very different from that of the housekeeping genes. Denamur et al. (28) found evidence that horizontal gene transfer of MMR genes may play an important role by increasing the rates of reacquisition of MMR function over those expected from compensating mutations only as implemented in our model.

The results presented here are robust – they are reproduced in multiple evolutionary runs. (see Supplementary Figures 4-7). A possible variation between runs is that in some cases the mutator phenotype remains fixed in the adapted population, again in agreement with experimental observations (*3, 15*).

Our microscopic evolutionary model of asexual organisms is simple and minimalistic. The unique feature of this approach, in contrast to traditional population genetics studies of mutation rates (1, 29, 30), is that it is based on first principles. That is, the effect of mutations is derived directly from genome sequences using a biophysically realistic model of protein stability and interactions in cytoplasm rather than postulated a'priori. As such, this model provides a description of physical principles of adaptation on all scales, from individual proteins to their assemblies in cytoplasm to populations of asexual organisms. On the population level, we found that adaptation always proceeds through transient fixation of a mutator phenotype. This is realized, on the microscopic level of proteins and their interactions, through a sequence of events which involve a peculiar interplay of intrinsic noise and genomic variation. The fact that a minimalistic ''first principles'' model was able to describe realistically many key aspects of molecular and cellular mechanisms of adaptation in real bacteria shows how evolution uses general physics as its ''design scaffold'', around which it builds a beautiful structure of living cells.

**Model and Methods**

In our model, organisms carry 5 genes whose sequences and structures are explicitly represented. Each gene has 81 nucleic acids. Once it is expressed into protein, it folds into a 3x3x3 compact lattice structure. We reduce the range of all possible 3x3x3 lattice structures, which totals 103,346, to randomly chosen 10,000 structures for faster calculation. $P_{nat}$ is the Boltzmann probability that the protein stays in its native structure whose energy is the lowest out of all 10,000 structures. There exist 144 rigid docking modes between two 3x3x3 lattice proteins, considering 6 surfaces for each protein and 4 rotations for each surface pair of two proteins (6x6x4). $P_{\text{int}}^{ij}$ is the probability that two proteins $i$ and $j$ form a stable dimeric complex in the correct docking mode. $P_{nat}$ and $P_{int}^{ij}$ are proportional to the Boltzmann weight factors of the native structure energy, $E_0$, and the lowest binding energy, $E_0^{ij}$ as follows:

$$P_{nat} = \frac{\exp[-E_0/T]}{\sum_{i=1}^{10000} \exp[-E_i/T]}, \quad P_{int}^{ij} = \frac{\exp[-f \cdot E_0^{ij}/T]}{\sum_{k=1}^{144} \exp[-f \cdot E_k^{ij}/T]}. \quad (2)$$

The binding constants $K_{ij}$ between proteins $i$ and $j$ are calculated as follows:

$$K_{ij} = \frac{1}{\sum_{k=1}^{144} \exp[-f \bullet E_k^{ij}/T]}, \quad (3)$$

and these values are substituted into the Law of Mass Action (LMA) equations in Eqs. S4 and S5 to determine the free concentrations of proteins $F_i$ and concentrations of their complexes $F_{ij}$. We use the Miyazawa-Jernigan pairwise contact potential for both protein structural and interaction energies (31), but scale protein-protein interactions by a constant factor, $f = 1.5$. We report environmental temperature $T$ (see text) in Miyazawa-Jernigan potential dimensionless energy units.

Simulations start from a population of 500 identical organisms (cells) each carrying 5 genes with initial sequences designed to be stable in their (randomly chosen) native conformations with $P_{nat} > 0.6$. At each time step, a cell can divide with probability $b$ given by Eq. 1. A division produces two daughter cells, whose genomes are identical to that of mother cells apart from mutations which occur upon replication with probability (mutation rate) $m$ per gene. If any protein in the cell loses its stability ($P_{nat} < 0.6$) by mutation, the cell is discarded. The death rate, $d$, of cells is fixed to 0.005/time unit, and the parameter $b_0$ is adjusted to set the initial birth rate to the fixed death rate ($b=d$). The control coefficient $\alpha$ in Eq. 1 is set to 100.

We simulated a chemostat regime: when the population size exceeded 5000 organisms, the excess organisms were randomly culled to bring the total population size to 5000. Initially expression levels are set equally for each protein at $C_i = 0.1$. The expression levels $C_i$ are inherited but they can fluctuate (implicitly modeling phenotypic changes and also mutations of transcription factor (TF) proteins and regulatory regions)—at each time step the value of $C_i$ can stay unchanged with probability $1-r$ or, with probability $r$, change. The magnitude of the change is random:

$$C_i^{new} = C_i^{old}(1+\varepsilon), \quad (4)$$

where $C_i^{old}$ and $C_i^{new}$ are the old and new expression levels of protein product of $i$-th gene, $\varepsilon$ is drawn from Gaussian distribution whose mean and standard deviation are 0 and 0.1, respectively. Paameter $r$ characterizes frequency of the fluctuation of protein copy numbers, we take $r = 0.01$ unless otherwise is noted. All molecular properties of individual proteins and their interactions are determined directly from genome sequences.

The concentration of free proteins $F_i$ is determined from the LMA:

$$F_i = \frac{C_i}{1 + \sum_{j=1}^{5} \frac{F_j}{K_{ij}}} \quad \text{for } i = 1,2,3,4,5 \tag{5}$$

where $K_{ij}$ is the binding constant of interactions between protein $i$ and protein $j$ (19) and concentrations of binary complexes between all proteins are given by the LMA relations:

$$F_{ij} = \frac{F_i F_j}{K_{ij}} \tag{6}$$

We determined, after each change (a mutation or a fluctuation in $C_i$), all necessary quantities by solving the LMA Eqs. (5) and (6) to find $F_1$, $F_{23}$, and $F_{45}$ evaluate the new $P_{nat}$ for mutated protein(s) and $P_{int}^{23}$ and $P_{int}^{45}$ for the complex of protein pairs to be in their specific binding conformation as explained above. Due to the nonlinear feature of the coupled LMA equations, we use an iterative method here. Once $C_i$ changes or a mutation occurs, the old set of $F_i$ is substituted into the right side of Eq. 5 and a new set of $F_i$ is calculated. This procedure iterates until the difference between old and new values of $F_i$ drops below the criteria of 1% of the new value.

To simulate a variable mutation rate, we take proteins 4 and 5 to be the components of DNA mismatch repair (MMR) machinery. We use the concentration of heterodimers between proteins 4 and 5 in their optimal docking mode—$G_{45} = F_{45} \cdot P_{int}^{45}$—as a metric to determine each organism's mutation rate. We assume a simple discrete model of mutation rate. If the $G_{45}$ of an organism is above a threshold $G_{45}^c$, then the organism has a low, wild-type mutation rate $m = 0.0001$ because it has sufficient functional error-correction complexes. If it is below $G_{45}^c$, then the organism transitions into a mutator phenotype with $m = 0.05$. In order to start simulations with organisms whose initial state is non-mutator, we designed the initial sequences for proteins 4 and 5 with $P_{nat} > 0.6$ and $P_{int}^{45} = 0.47$, using design algorithms described in (32, 33) and set the threshold $G_{45}^c$ to $0.6 \cdot G_{45}^0$, where $G_{45}^0$ is the initial concentration of $G_{45}$. We do not deign initially interactions between products of genes 1-3, so that populations start from non-adapted growth rate conditions.

*Acknowledgements.* We thank Sergei Masolv for comments on the manuscript. This work is supported by the NIH.

# Supplementary Information

**The relationship between the level of fitness and the frequency of mutators**

In the previous section of the distribution of the relative fitness change upon 1,000 single mutations, we observed that the mutator allele could get fixed regardless of the perturbation of fitness landscape by a new environment if the population growth was limited. Here, we elucidated how the level of fitness affects the fractional population of mutators. Both Chao's (1) and de Visser's (2) experiments seem to suggest that frequency of mutators may be related to the level of fitness. Our chemostat regime simulates a limited-resource environment in which the population is randomly culled the environment's carrying capacity if the population size exceeds this limit. High fitness in such an environment causes greater production of new organisms and hence a larger excess of organisms over the carrying capacity. Thus, more organisms must be culled at high fitness for each unit of time, which means a faster speed of random drift. Individual mutators that arise in the system would face a greater danger of elimination through random culling. To test this hypothesis, we designed control simulations with fixed fitness. All effects of mutations on fitness were neutralized by fixing the growth rate at a constant instead of deriving it from genomic sequences according to Eq.1 of the main text. However, we still left a protein structural constraint which removes organisms due to protein malfunction if any of its proteins lost stability ($P_{nat} < 0.6$). This constraint provided a weak selection against deleterious mutations which arose more frequently in mutator clones. Four independent simulations were carried out for five fixed birth rates, $b$=0.01, 0.05, 0.1, 0.5, and 1.0 until $t$=10,000. We waited 8,000 time steps to equilibrate the system, and 81 time points were sampled from t=8,000 to 10,000 every 25 time steps for each pathway, which totaled 81x4=324 time points for each fixed birth rate. Figure S5 shows that the frequency of mutators dropped when the growth rate increased. This trend means that the level of fitness determines the frequency of mutator clones through random drift, because mutators are supplied at a constant rate by stochastic switching. The frequency decayed up to $b$=0.1 in the figure. If the growth rate is higher than 0.1, the frequency seemed to reach a baseline caused by the influx of mutator by stochastic switching. In this regime, all mutators could be discarded by random drift every generation due to the high growth rate, and the residual frequency might be just a recurring supply of mutator clones from non-mutators.

**Supplementary Figures**

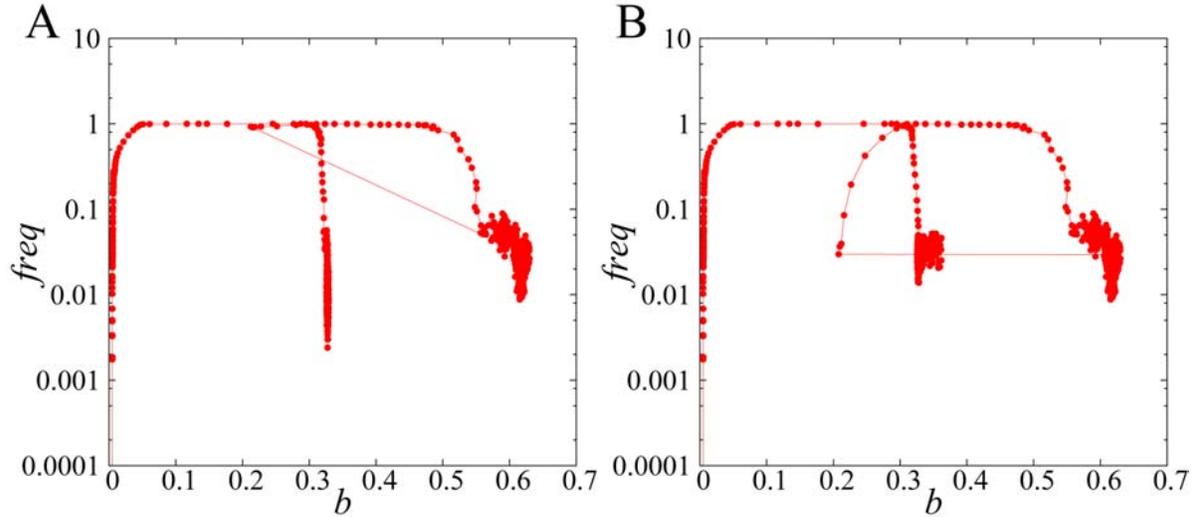

**Figure S1. Fitness vs. mutator frequency.** Traces of the evolutionary pathway of the population in Fig 2B (A) and in Fig 2C (B) are presented in terms of fitness (*b*) versus the frequency of mutators (*freq*). Both traces started from the left bottom corner of the plots. The frequency reached 1 as fitness approached *b*=0.05 from *b*=0.005 and stayed until *b*=0.5. After fitness exceeded 0.5, the frequency of mutators started to drop. Both temperature jump (A) and simulated starvation (B) decreased the growth rate and shifted the traces sharply back to the center of the plots. In A, mutators arise directly from temperature destabilization as demonstrated by the diagonal line. However in B, the corresponding line is flat, indicating that mutator fixation followed fitness drop as a separate event. In both cases, the mutator frequencies dropped again when fitness exceeded 0.3.

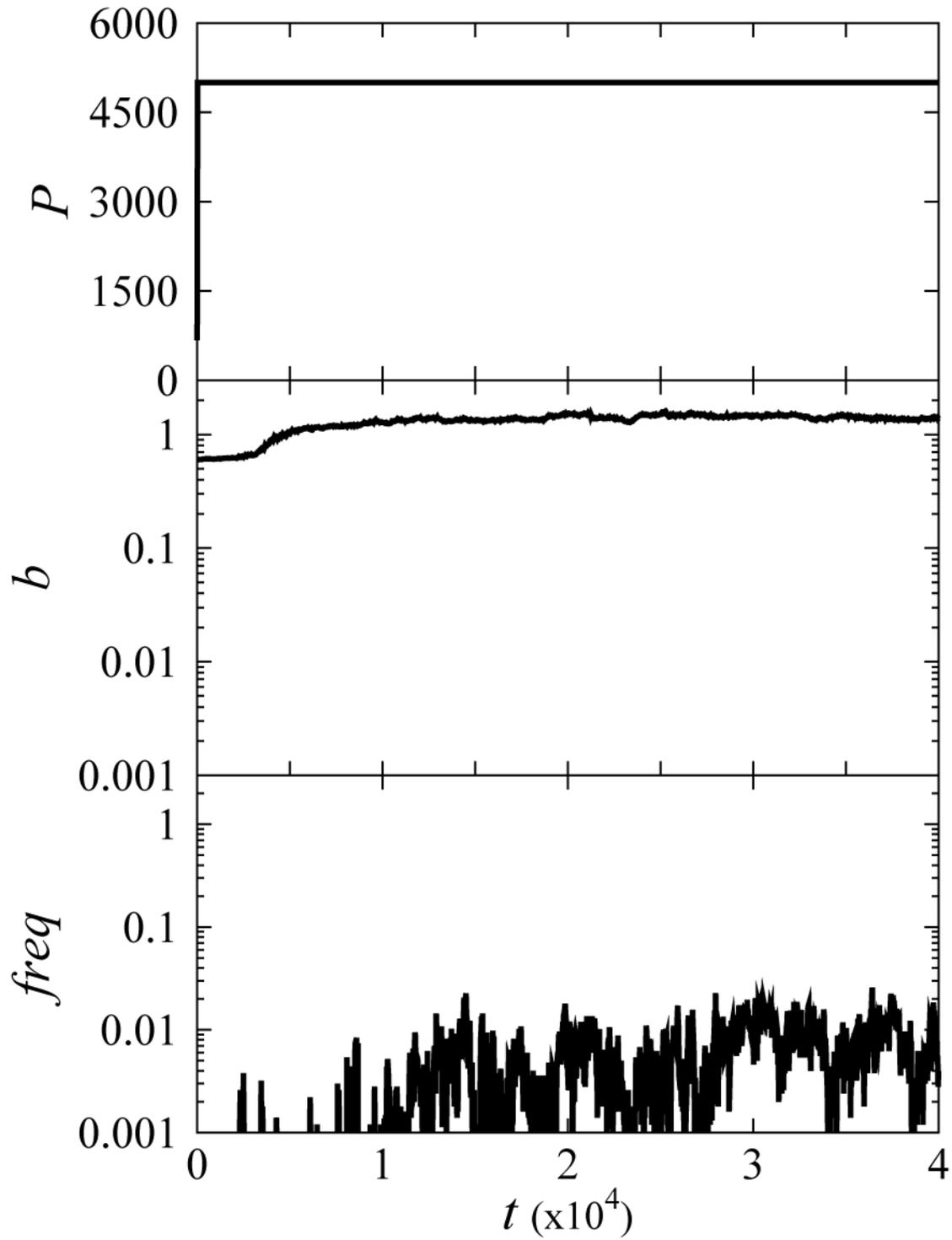

**Figure S2. Population dynamics with initial high fitness.** Panels show population ($P$), mean birth rate ($b$), and frequency of mutators in the population (*freq*). If the simulations start at a high initial birth rate ($b=0.6$), no preference is found for the mutator phenotype and mutators are never fixed. These simulations use the same pseudorandom number generator seed as the simulations in Figure 2.

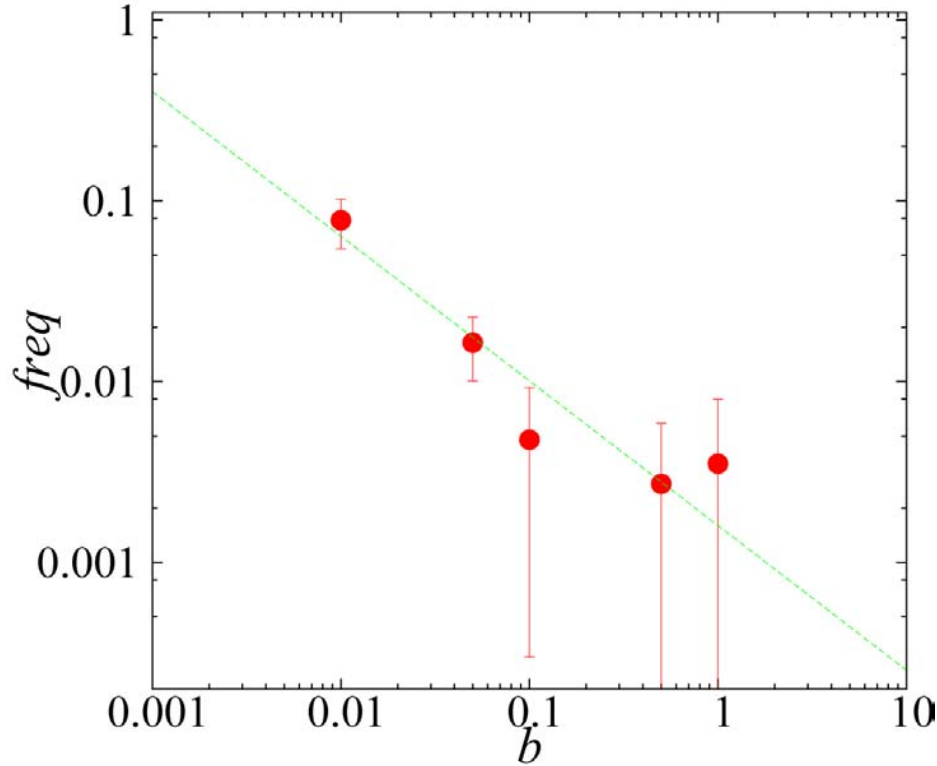

**Figure S3. Analysis of random drift through control simulations with constant fitness.** The frequency of mutators (*freq*) is plotted against constant fitness (*b*) on a log-log scale. In this control simulation the organism duplication rate *b* is kept constant rather than determined by Eq.1 of the main text. Each point represents mean frequency of mutators at various constant fitness. Because the effects of mutations on fitness are neutralized by constant fitness, mutator frequency did not depend on hitchhiking and was completely determined by the amount of random genetic drift that corresponds to each level of fitness. Stochastic phenotype switching supplied mutator clones to the population at a constant rate depending on gene expression level fluctuation rate, *r*. Higher fitness produced more excess organisms, which were eliminated due to the limited capacity of a finite population. In turn, initially rare mutator clones faced a greater danger to be purged away from the population because of their higher death rate due to mutations causing instability of their proteins (weak selection against lethal mutations decreasing $P_{nat}$ below 0.6), and thus the frequency of mutators dropped as the growth rate increased.

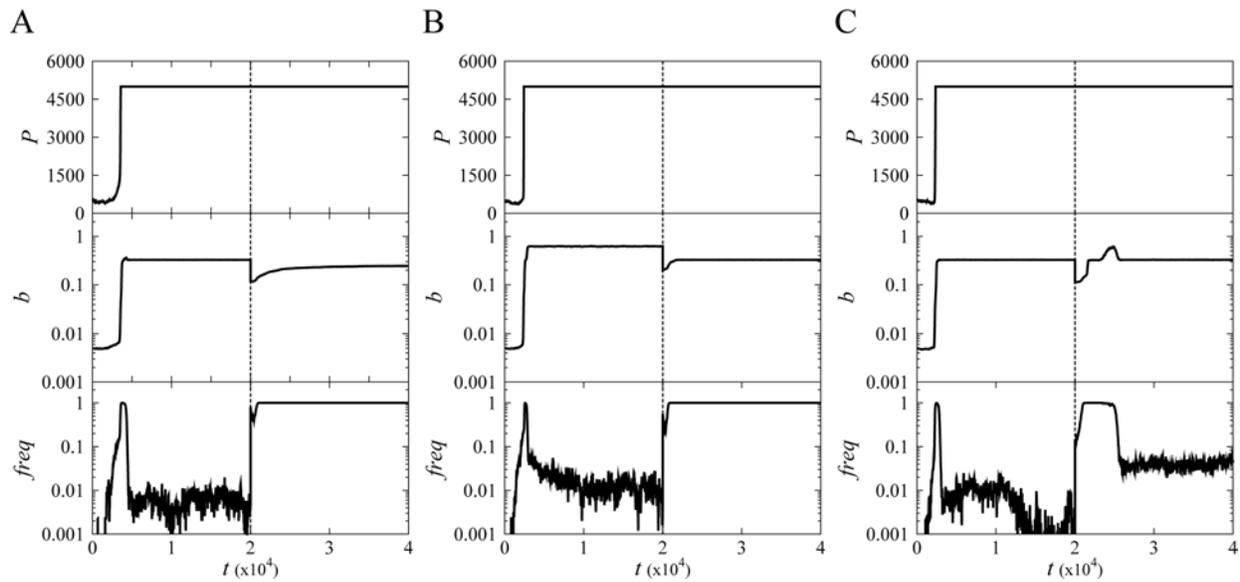

**Figure S4. Population dynamics of the temperature jump simulation: additional runs.** Panels show population (*P*), mean birth rate (*b*), and frequency of mutator allele (*freq*) in the population as function of time (*t*). Three more simulations (A, B, and C) that mimicked the temperature jump like Fig. 1B were performed independently. The blue line in each panel at *t*=20000 marks the time of temperature increase. The temperature jump caused a drop in *b* and another mutator fixation event that allowed the population to find a beneficial mutation that enhanced birth rate. After the second fixation, mutators in panels (A) and (B) remained persistent to the end of simulation. But non-mutators in (C) were preferred like Fig. 1B after adaptation.

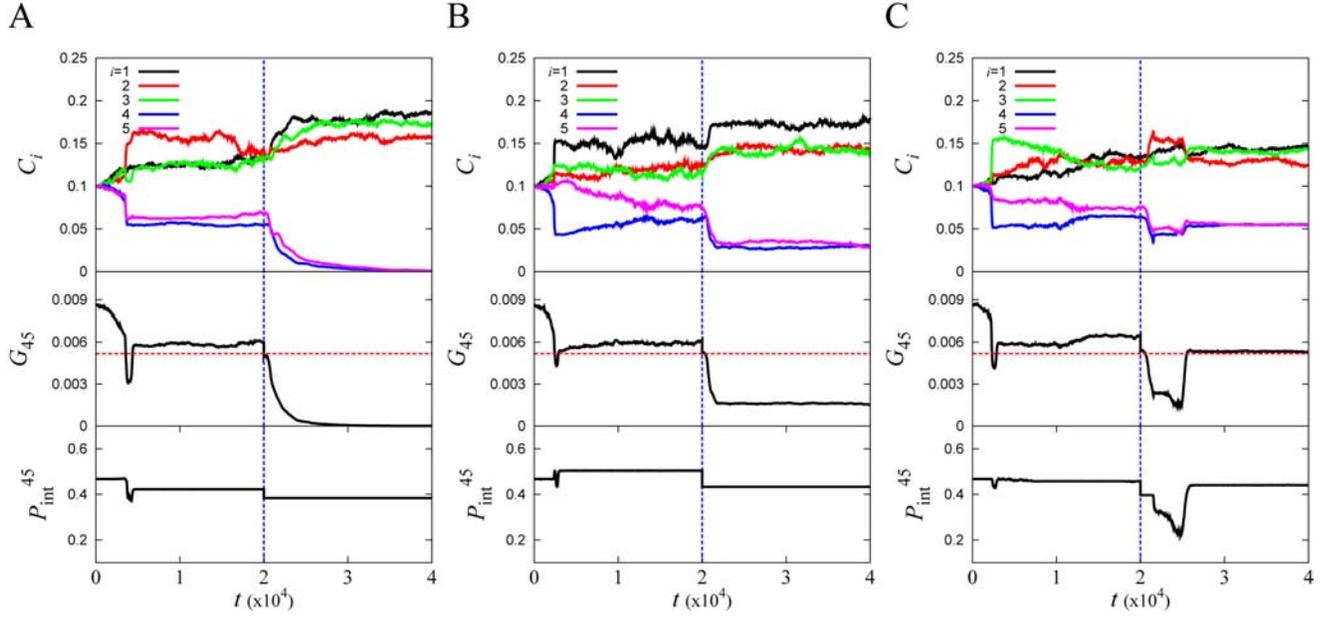

**Figure S5. Microscopic variables of population dynamics of the temperature jump simulation: additional runs.** Panels presents mean concentrations ($C_i$) of each protein product, mean heterodimer concentration of proteins 4 and 5 which constitute the functional MMR complex ($G_{45} = F_{45} \cdot P_{int}^{45}$), and mean heterodimer stability of the functional complex ($P_{int}^{45}$) as function of time ($t$). Each panel of (A), (B), and (C) respectively represents microscopic variables corresponding the simulation of panel (A), (B), and (C) in Fig. S6. The red line in the $G_{45}$ panel is set at $G_{45}^c$, the critical heterodimer concentration below which an organism becomes a mutator. Mutator clones at $t$=20000 were induced by loss of $P_{int}^{45}$ by temperature jump. Mutators in (A) and (B) persisted, because they could not produce mutations recovering the MMR system. But mutators in (C) disappeared after adaptation as new mutations restored the MMR system by enhancing $P_{45}$.

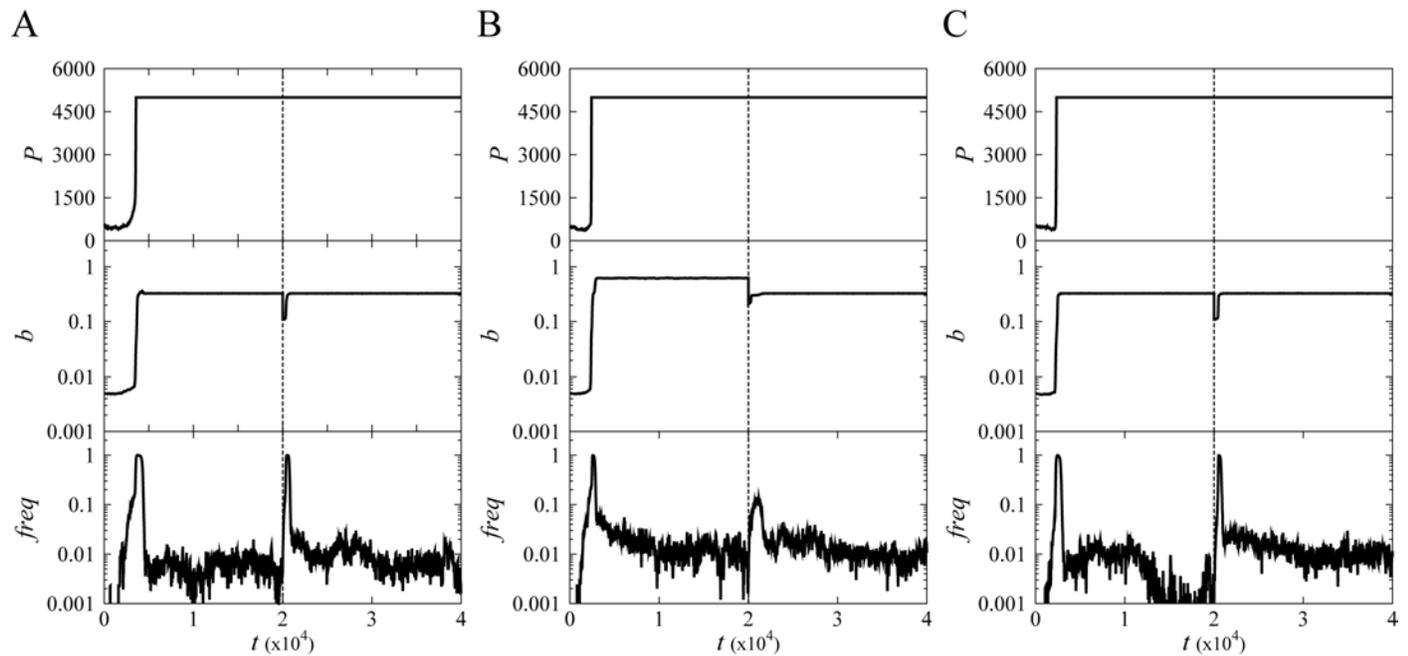

**Figure S6. Population dynamics of the starvation simulation: additional runs.** Panels show population ($P$), mean birth rate ($b$), and frequency of mutator allele (*freq*) in the population as function of time ($t$). Three more simulations (A, B, and C) that introduced the starvation condition like Fig. 1C were performed independently. The blue line in each panel at $t=20000$ marks the time of fitness drop due to starvation. The starvation caused 100% fixation of mutators for (A) and (C). Mutators in (B) could not get fixed (the maximum frequency of mutators is ~0.1), because the population improved fitness too quickly for mutators to get fixed. No fixation of mutators in (B) resulted in lower level of fitness. After adaption, non-mutators werepreferred.

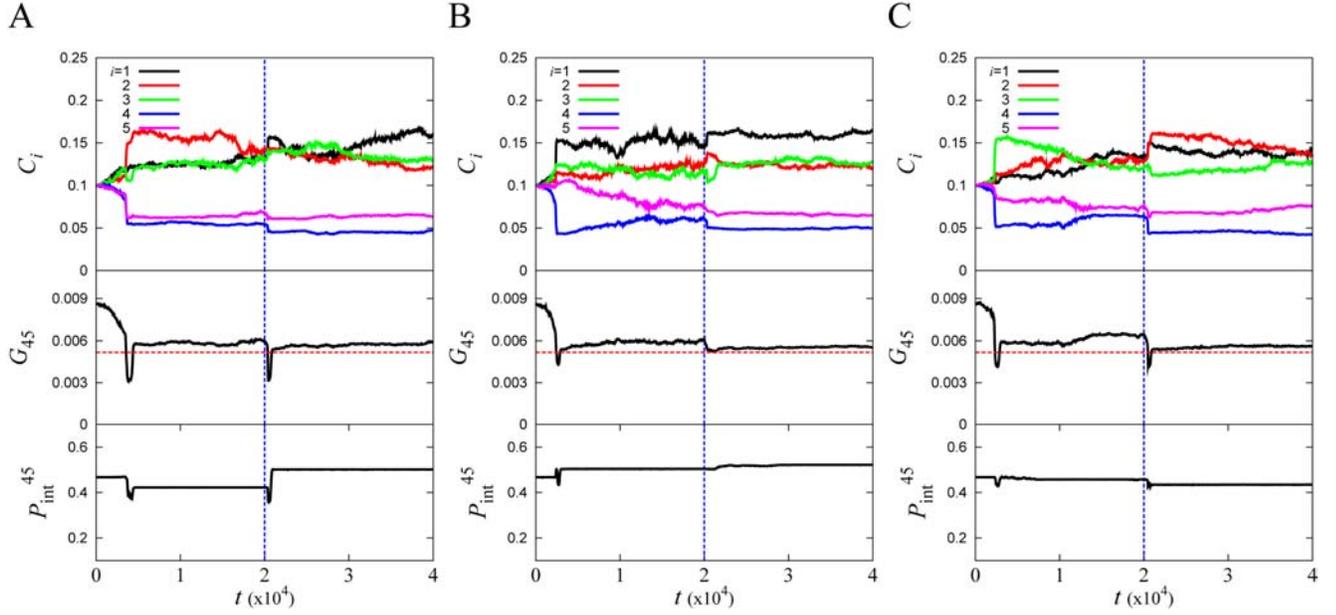

**Figure S7. Microscopic variables of population dynamics of the starvation simulation: additional runs.** Panels presents mean concentrations ($C_i$) of each protein product, mean heterodimer concentration of proteins 4 and 5 which constitute the functional MMR complex ($G_{45} = F_{45} \cdot P_{int}^{45}$), and mean heterodimer stability of the functional complex ($P_{int}^{45}$) as function of time ($t$). Each panel of (A), (B), and (C) respectively represents microscopic variables corresponding the simulation of panel (A), (B), and (C) in Fig. S8. The red line in the $G_{45}$ panel is set at $G_{45}^c$, the critical heterodimer concentration below which an organism becomes a mutator. Mutator clones at $t$=20000 in (A) and (C) were induced by stochastic switching and as they generated mutations disrupting the binding strength of the MMR complex, they became fixed. However, mutators in (B) could not generate those mutations, and the frequency of mutators only reached ~0.1 because their mutator phenotypes could be easily switched off by the fluctuation in gene expression. After adaptation, mutators in (A) and (C) disappeared due to new mutations restoring the MMR complex by increasing $P_{int}^{45}$, even though the increment of $P_{int}^{45}$ was very small in (C).